\documentclass[12pt,preprint]{aastex}
\input epsf.sty

\shorttitle{Magnetic field variability in $\gamma $ Equ}
\shortauthors{V.D. Bychkov, L.V. Bychkova\inst{1}, and J. Madej }

\begin{document}

   \title{ Investigation of the rapid magnetic field variability \\
          in $\gamma $ Equ}

\author{V.D. Bychkov\altaffilmark{1}, L.V. Bychkova\altaffilmark{1}, 
        and J. Madej\altaffilmark{2} }
\altaffiltext{1}{Special Astrophysical Observatory of the Russian   
       Academy of Sciences, Nizhnij Arkhyz, 369167 Russia }
\altaffiltext{2}{Warsaw University Observatory, Al. Ujazdowskie 4,
       00-478 Warsaw, Poland}

\email{vbych@sao.ru}

\begin{abstract}
We investigated magnetic variability of the roAp star 
$\gamma$ Equ (HD 201601) with high time resolution. Measurements of the
stellar longitudinal magnetic field $B_e$ were performed with the hydrogen
line polarimeter and the 6-m optical telescope of Special Astrophysical
Observatory (Russia) on 20/21 August 1989. 
We obtained a single 3-hour series of 1720 rapid $B_e$ measurements for this
star with the average time resolution of 6.6 sec. The averaged value
of $\langle B_e \rangle $ on this night equals $-750 \pm 22$ G.
Both the power spectrum and the amplitude spectrum of the $B_e$ time series
are essentially flat. However, they show the presence of a marginally 
significant magnetic period at $P_B = 3.596$ min. which is real with the
probability 67 \%. 
The amplitude of the magnetic field variations with this period equals 
$347 \pm 31$ G. We did not detect variations of $B_e$ field in $\gamma$ Equ
with the well-known photometric period, $P_{phot}=12.44$ min. 

\end{abstract}


\section{Introduction}

Rapid oscillations in some of Ap-stars were discovered by Kurtz \& Wegner
(1979), and Kurtz (1982). These oscillations are due to high-overtone p-mode
pulsations, and appear as rapid periodic photometric variations in B and V
filters of the Johnson broadband photometric system. Periods of 
oscillations are in the range 4--16 min., and photometric variations
exhibit low amplitudes. They are independent on photometric
variations caused by the stellar rotation.
  
Rapidly oscillating Ap-stars are often refered to as roAp-stars,
and this class presently contains 32 stars (Kurtz \& Martinez 2000).

Search for rapid variations of the radial velocity in $\gamma$ Equ (HD 201601,
A9p SrCrEu, $V=4.7$ mag) was initialized by Bychkov (1987, 1988). Paper by
Zverko et al. (1989) reported the search for a rapid variability of its 
magnetic field and radial velocity. All these early investigations used the
photometric technique of observations, and did not yield conclusive results
due to low accuracy. Recently, Kochukhov \& Ryabchikova (2001) performed
similar research using modern equipment, and have obtained reliable phase
variations of the radial velocity in $\gamma$ Equ.

Recently some research groups attempted to observe rapid variations
of its magnetic field. Particularly interesting is a possible variability
of the magnetic field with the pulsation period of $\gamma$ Equ. 
Leone \& Kurtz (2003) reported a discovery of rapid variations of the
stellar longitudinal (effective) magnetic field $B_e$ with a period of
12.1 min., and the amplitude equal to $240 \pm 37$ G. In a later communication
Savanov et al. (2003) have found variations of the magnetic field modulus
with the period 12.53 min. with the amplitude $99 \pm 53$ G. 
The authors reported in their paper variability of a separation between
the resolved Zeeman components of the Fe II 6149.2 A line. 

On the other hand, Kochukhov et al. (2004a) reported the absence of variations
of the longitudinal magnetic field $B_e$ in $\gamma$ Equ, which directly
contradicts earlier claims by Leone \& Kurtz (2003). Kochukhov et al. (2004a)
obtained a series of 210 spectropolarimetric observations with a high time 
resolution, and found no $B_e$ variations with amplitudes above $40-60$
G in the circularly polarized components of 13 Nd III lines.  Such a limit 
represents their $3\sigma$ confidence level.
Moreover, Kochukhov et al. (2004b) have shown that possible variations of
the surface magnetic field in $\gamma$ Equ do not exceed $5-10$ G in short
time scales (observations in Fe II 6149.25 and Fe I 6173.34 lines).

Recently Hubrig et al. (2004) concluded a search for rapid variability of
the longitudinal $B_e$ field in several magnetic stars (observations in 
circularly polarized profiles of hydrogen lines). Unfortunately, their
results on $\gamma $ Equ are not conclusive due to a small number of $B_e$
points.

This paper presents results of our search for rapid periodic 
variations of the longitudinal $B_e$ field in $\gamma$ Equ. We have 
performed Fourier spectral analysis of archival time series of 1720 $B_e$
continuous measurements in the search for periods. 
Individual $B_e$ points were measured with a very fine time resolution
($\approx$ 6 seconds on the average).

\section{ Observations and data analysis }

\subsection{Instrumentation}

We used the hydrogen line polarimeter (Bychkov et al. 1988; Shtol' 1991, 1993)
for measurements of the longitudinal stellar magnetic field $B_e$.
The polarimeter was placed in the prime focus of the 6-meter telescope in
order to minimize the influence of instrumental effects from other optical
elements of the telescope. 

The hydrogen line polarimeter was used to determine circular polarization
of radiation ($V/I$) simultaneously in both wings of a hydrogen line.
Actual observations were performed in two Balmer lines, $H_\beta$ and
$H_\gamma$. The polarimeter cuts out two narrow bands from the line wings
of the width 18 {\AA } each, using a special mask. Both spectral 
transmission bands were located symmetrically about the line core, and
the edges of transmission were shifted by 2 {\AA } away from the core. 

Profiles of both $H_\beta$ and $H_\gamma$ lines are usually very similar
but are not identical.

First, we determined the average profiles of both lines separately with
the same mask, which was described above. Then the averaged line profiles
were used to determine the transformation coefficient between the degree
of circular polarization measured in line wings, and the stellar effective
magnetic field $B_e$. That transformation was measured in both $H_\beta$
and $H_\gamma$ line wings simultaneously, taking into account a real
contribution of each line to the total flux (weighted by the number of 
photons received per second). At the time of actual observations of 
$\gamma$ Equ the transformation coefficient was equal to $K_b=37000 $ G 
per 1~\% of the circular polarization.

Measurements of the circular polarization were performed both in the blue
and red line wings simultaneously, which allowed us to find and take into
account instrumental contribution to the observed polarization. 
Stability of the modulator and other optical and electronic components
of our instrumentation were constantly monitored at the time of observations.
Moreover, at the time of the observing run we measured also other stars with
the well known values of the magnetic field $B_e$, usually 2-3 times per run.

\subsection{Observations}

Observations of the magnetic field $B_e$ in $\gamma$ Equ were performed
during a single night of 20/21 August 1989. Heliocentric moments of the
beginning and the end of our observations are JD 2447759.414471 and .545842,
respectively.

Time length of a single exposure was determined by the time of accumulation
of up to 800000 photons, and it varied between 3 and 24 seconds. The average
time length of a single observation was equal to 6.6 sec in our run.
Time span between the end of a previous exposure and the beginning of the 
following one did not exceed 0.05 sec, which means that a series of 
measurements was practically continuous. The total time length of our 
observing run of $\gamma$ Equ was equal to 189 minutes, 
which enabled us to collect 1720 estimates of $B_e$ of this star.
Such a large number of $B_e$ points in a single, uninterrupted series
allowed us to compute reliable power spectra of the above data.

Photon statistics implies some rough estimates of the error $\sigma$ of a
single magnetic field determination. The error
\begin{equation}
  \sigma \approx K_b \times 100\% / \sqrt{2 N} = 2925 \, {\rm G} \, ,
\label{equ:sigma1}
\end{equation}
where $N$ denotes the number of accumulated photons (Shtol' 1993).

The error of a single $B_e$ measurement can be estimated in an independent
way. We assume that the projected magnetic field $B_e$ does not change at
the time of observations, and the scatter of individual $B_{ei}$ points
results from random observational errors. Then, the error of a single
measurement equals
\begin{equation}
  \sigma = {1\over {n-1}} \left[ \sum_i (\langle B_e \rangle - B_{ei})^2 
     \right]^{1/2} = 2837 \, {\rm G} \, , 
\label{equ:sigma2}
\end{equation}
for the actual series of $n=1720$ measurements. One can see that both
independent estimates of the error of an individual $B_e$ observation differ
rather insignificantly.

We stress here that we have presented in this paper the first magnetic field 
measurements obtained with such a fine time resolution, without any significant
readout time between measurements.

\subsection{Spectral analysis of a time series}

\begin{figure}
\resizebox{\hsize}{0.6\hsize}{\rotatebox{0}{\includegraphics{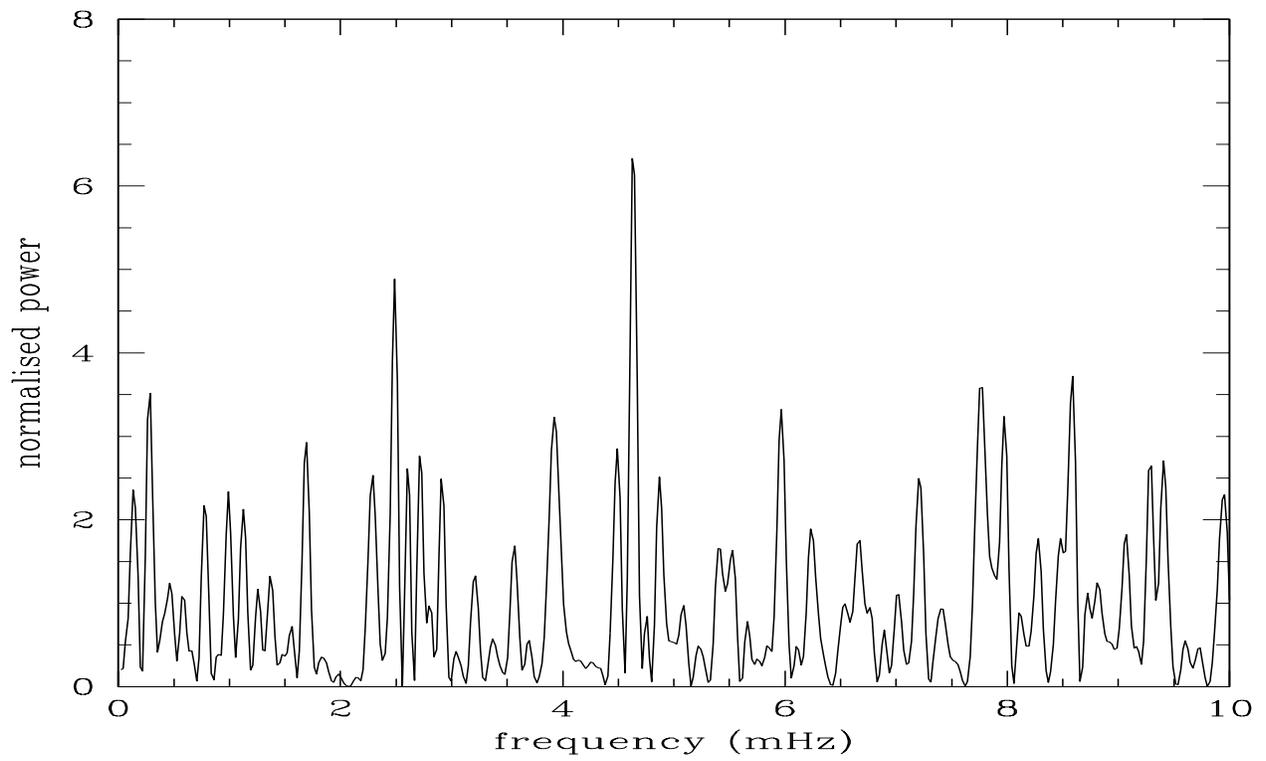}}}
\caption[]{Power spectrum of the series of 1720 $B_e$ observations. }
\label{fig:power}
\end{figure}

We have performed spectral analysis of the above $B_e$ time series using
two different numerical techniques: the power spectrum and
the amplitude spectrum analysis. First, the 
standard Fast Fourier Transform subroutines were used to analyse the
above series of 1720 $B_e$ measurements of $\gamma$ Equ. We have computed
its power spectrum in the frequency range $0.185-16.67 \, {\rm mHz}$, which
corresponds to periods ranging from 90 min. to 1 min. Fig.~\ref{fig:power}
shows the power spectrum, which is rather noisy in this frequency range.
We have identified the distinct peak at 4.635 mHz,
significantly exceeding the level of a spectral noise. 
The peak frequency corresponds to the period $P_B = 3.596$ min.

Numbers on the vertical axis of Fig.~\ref{fig:power} denote the values of the
normalised power of our $B_e$ series. Spectral power of a time series is
the square of the Fourier transform. In case of magnetic $B_e$ observations
power can be measured in G$^2$/mHz$^2$. Normalised power of a signal is 
defined as the ratio of the true power and its variance, and therefore it
is a dimensionless variable (Horne \& Baliunas 1986).

The advantage of using the normalised power is that Scargle (1982)
proposed a relation between the power at a spectral peak and the probability
that the peak corresponds either to real or false period in a time series
(False Alarm Probability test).

The value of a normalised power at the peak of 4.635 mHz in our $B_e$
measurements equals 6.327. False Alarm Probability test of the latter paper
predicts, that the peak is real with the probability 67 \%
(cf. Section II.b in Horne \& Baliunas 1986). Therefore, the spectral
peak is a 1 sigma event and the corresponding period is not necessarily
a real feature.


\begin{figure}
\resizebox{\hsize}{0.6\hsize}{\rotatebox{0}{\includegraphics{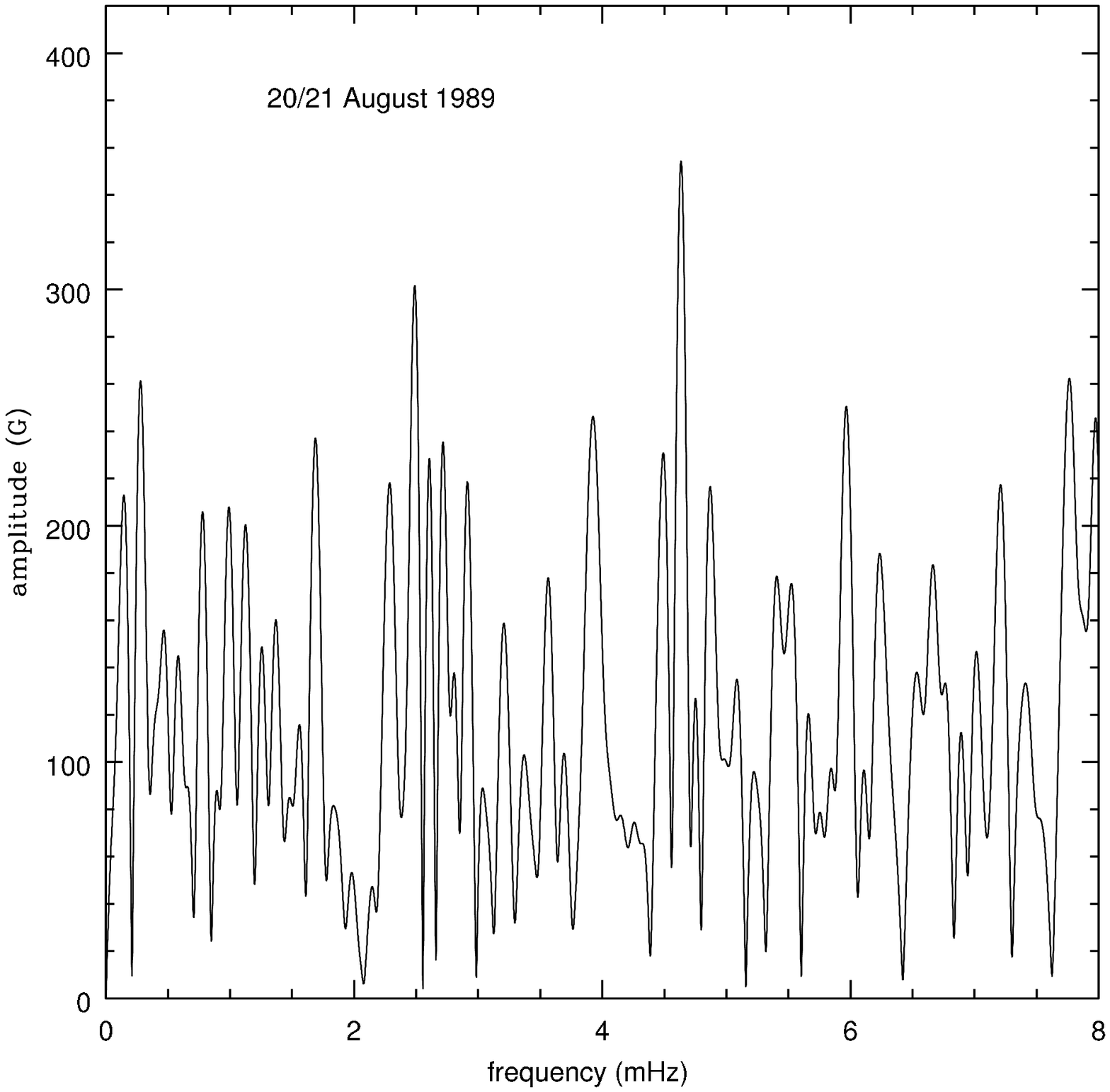}}}
\caption[]{The amplitude spectrum for our series of 1720 $B_e$ 
   measurements. }
\label{fig:amplitude}
\end{figure}

Subsequently, we have analysed the amplitude spectrum of our $B_e$ time series
with the method developed by Kurtz (1985) and with his software (Kurtz 2004).
Fig.~\ref{fig:amplitude} shows the amplitude spectrum which is also flat
with a much lower amplitude peak at the same frequency. 
Fig.~\ref{fig:amplitude}
allows one to estimate the noise level of our data, which is of the order of
250 G. The amplitude of possible periodic $B_e$ variations shown there is of
the order of 350 G. 

\begin{figure}
\resizebox{\hsize}{0.7\hsize}{\rotatebox{0}{\includegraphics{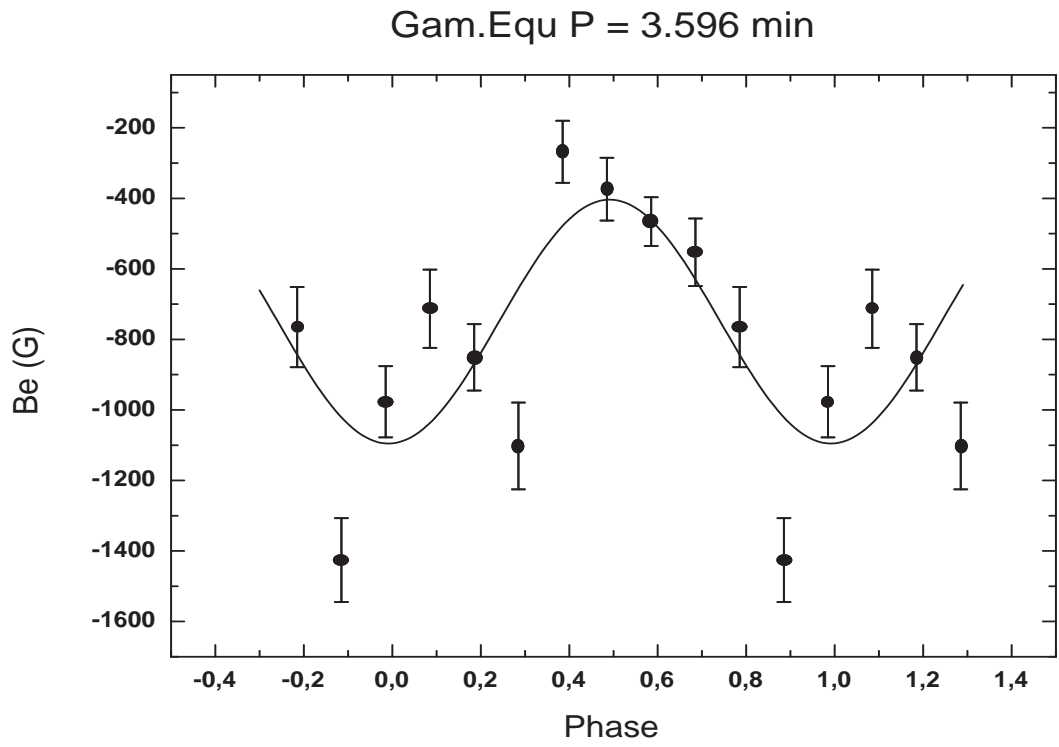}}}
\caption[]{Magnetic sine phase curve best fitted to our 
$\langle B_e \rangle $ points computed for the period $P_B=3.596$ min. }
\label{fig:p3}
\end{figure}

Fig.~\ref{fig:p3} presents the magnetic phase curve for $\gamma$ Equ
corresponding to the period $P_B = 3.596$ min. 
Phase curve was fitted to ten $\langle B_e \rangle $ points, which
were obtained by averaging of individual noisy $B_e$ measurements in
phase bins of the width $0.1 \, P_B$.
The value of $\langle B_e \rangle $ averaged over all phases equals
$-750 \pm 22$ G, and the amplitude of the phase curve is $ 347\pm 31$ G.

One should note that the phase plot in Fig.~\ref{fig:p3} does not 
confirm the reality of the peak in Fourier transform.
Any high peak in the FT will produce a phase plot where you can see
apparent variation. That is true whether the peak is real or spurious
(Kurtz 2004). The only useful information given by Fig.~\ref{fig:p3}
is that the phase coverage of the $B_e$ data in our series is
satisfactory, assuming that the $P_B = 3.596$ min period, or rather
4.635 mHz peak frequency, is real.

\section{ Discusion}

Most of the previously analysed magnetic measurements of $\gamma$ Equ
were obtained in metal lines. Consequently, the results could be affected
by a nonuniform distribution of these elements over the stellar surface.
Such a well known effect is common among Ap stars. Moreover, $B_e$ 
measurements in metal lines in principle may be distorted by other effects,
e.g. by complex blending of lines, cross-over effect, etc. 

The only exception is the paper by Hubrig et al. (2004) who obtained their
$B_e$ measurements in wings of the hydrogen Balmer lines. As a consequence,
their $B_e$ data are free of the above effects, and they seem to be more
credible since the hydrogen line $B_e$ data represent best the value of 
the longitudinal component of the global magnetic field integrated
over the stellar disc. 

It is a well known fact, that the profiles and equivalent widths of the
Balmer lines exhibit periodic variations in most of Ap stars. Such variations
are the result of stellar rotation, and proceed with the corresponding
rotational period. For example, observations of the periodic variability of
$H_\beta$ line with $P_{rot}$ were presented by Madej (1983) and Musielok
\& Madej (1988) for a group of bright Ap stars. The presence of rapid
variations in the wings of Balmer lines in $\gamma$ Equ cannot be excluded,
since momentary line profiles should reflect all changes in the structure of
a stellar atmosphere caused by pulsations. However, the impact of a 
possible rapid variability of Balmer line profiles on the $B_e$ 
determinations is unknown and will be the subject of our research in future.

One can also note, that the amplitudes of magnetic curves observed in 
hydrogen lines can be distinctly larger than the amplitudes measured in 
metal lines. Such an effect arises since (in atmospheres of magnetic Ap stars)
far wings of hydrogen lines are formed mostly in much deeper layers than
metal lines. Therefore, if strength or direction of the local surface magnetic
field depends on depth, i.e. if it exhibits nonzero gradient, then the 
apparent intensity of $B_e$ and its temporal variations can differ depending
on the depth of a spectral line formation.

Hubrig et al. (2004) presented only 18 individual measurements of the 
longitudinal magnetic field $B_e$ of $\gamma$ Equ, due to the cloudy weather
during their 2.5 hour observing run. They did not find any significant
traces of the period. In particular, Hubrig et
al. (2004) did not confirm any periodic variability close to the four 
pulsation periods in the range 11.68--12.44 min. (Martinez et al. 1996).

We have taken 18 magnetic field $B_e$ points from Hubrig et al. (2004),
and searched for a period $P$ disregarding small number of available points.
They show pure spectral noise with amplitude of $\approx 150 $ G in the 
frequency range $0-6$ mHz, hence they do not confirm our marginal period
of $P_B $.

We cannot rule out the possibility, that processes which cause rapid
pulsations in $\gamma$ Equ are not stable, and cause a drift of pulsation 
periods and amplitudes with time. One can speculate, that our observations
and the period of $B_e$ obtained in 1989 ($P_B=3.596$ min.) do not 
correspond to $B_e$ periods obtained at a later time. 

Such a hypothesis would explain, why Leone \& Kurtz (2003), and Savanov
et al. (2003) could obtain a period close to the photometric period
$P_{phot}=12.44$ min. On the other hand, Kochukhov et al. (2004a,b) might
have observed $\gamma$ Equ during a time interval, when the amplitude of
rapid variations of $B_e$ decreased below the detection limit.

Kochukhov et al. (2004a,b) measured magnetic field $B_e$ in lines of Fe I,
Fe II, and Nd III, which may not be very sensitive to the
magnetic field variations (Nd III lines were useful to measure variations
of radial velocity). Plachinda \& Polosukhina (1994) showed, that values
of $B_e$ determined in different lines can differ by a factor up to a few
times. Therefore, we propose that the most representative values of $B_e$
are those measured in wings of hydrogen Balmer lines.

\section{Summary }

In this paper we present our high time resolution observations of the 
global longitudinal magnetic field in the Ap star $\gamma$~Equ = HD 201601.
The series of 1720 measurements was obtained during 189 min. of a single 
uninterrupted run on August 20-21, 1989. We found that the average value of
$\langle B_e \rangle $ at the time of observations equals $-750 \pm 22$ G.

Spectral analysis of this time series has shown that the frequency spectrum
of $\gamma$ Equ is flat. However, we have found possible variations of the
global-scale $B_e$ with the period $P_B=3.596$ min. There is only a 67 \%
probability that this period is a real feature (1 sigma event), and
therefore, its significance is not high.

Both the power spectrum and the amplitude spectrum of our data reveal no
signal at the frequency of pulsations, which was identified with the 
well-known period of rapid photometric variations, $P_{phot}=12.44$ min.
Our amplitude spectrum (Fig.~\ref{fig:amplitude}) shows, that if rapid 
variations of $B_e$ with the period 12.44 min. existed in 1989,
then their amplitude was lower than $\approx$ 250 G (noise level).

There exists a number of theoretical papers, which attempt to explain the
existence of nonradial pulsations in Ap stars with very short periods
of the order $\sim$ 10 min (Bigot et al. 2000; Cunha \& Gough 2000;
Bigot \& Dziembowski 2002). We believe, that the model of inclined pulsator
presented in the latter paper is the most relevant one to discuss 
possible variations of the global magnetic field in $\gamma$ Equ with
periods as short as 3.6 min.

\begin{acknowledgements}

Our thanks are due to Don Kurtz for providing his Fortran software used
here to compute the amplitude spectrum of $\gamma$ Equ and for critical
comments regarding our results. We also thank Kazik St\c epie\'n
for careful reading of our paper and numerous suggestions. 
This paper was supported by grant No. 1 P03D 001 26
from the Polish Committee for Scientific Research. 

\end{acknowledgements}

\end{document}